\documentclass[10pt,aps,nofootinbib]{revtex4}
\usepackage{amsmath,graphics,epsfig}
\usepackage{color,hyperref}
\usepackage{datetime}
\usepackage[utf8]{inputenc}
\begin{document}

\title{\bf  Effective Field Theory of Majorana Dark Matter}
\author{Huayong Han,$^{a}$~Hongyan Wu$^{b,c}$~and Sibo Zheng$^{c}$ \\
{\small $^{a}$ CAS Key Laboratory of Theoretical Physics, Institute of Theoretical Physics, \\
Chinese Academy of Sciences, Beijing 100190, China}\\
{\small $^{b}$ School of Physics, Peking University, Beijing 100871, China}\\
{\small $^{c}$ Department of Physics, Chongqing University, Chongqing 401331, China}}

\begin{abstract}
We revisit thermal Majorana dark matter from the viewpoint of minimal effective field theory.
In this framework, analytic results for dark matter annihilation into standard model particles are derived. 
The dark matter parameter space subject to the latest LUX, PandaX-II and Xenon-1T limits 
is presented in a model-independent way. 
Applications to singlet-doublet and MSSM are presented.
\end{abstract}

\maketitle

\section{Introduction}
In the viewpoint of standard cosmology, 
cold dark matter (DM) is a neutral particle beyond the standard model (SM),
which has not been observed in either particle astrophysical or collider experiments.
Due to its electrically neutral property, it is natural to consider DM as a Majorana fermion.
Majorana DM appears in a lot of well known models such as neutralino \cite{Jungman:1995df}, 
singlet-doublet \cite{0510064, 0705.4493, 0706.0918, 1109.2604, 1505.03867,1311.5896,1603.07387,1707.030948},
Higgs-portal \cite{0611069,1005.5651,1112.3299,1203.2064} 
and $Z$-portal \cite{1708.00890,1612.07282,1402.6287,1609.09079,1611.05048,1506.06767} DM.

For Majorana DM, the effective Lagrangian at the weak scale is described by,
\begin{eqnarray}{\label{Lag}}
\mathcal{L}=\mathcal{L}_{\text{SM}}+\mathcal{L_{\text{dark}}}(\chi, \cdots),
\end{eqnarray}
where the SM Lagrangian $\mathcal{L_{\text{SM}}}$ contains interactions between DM mediators $h$ and $Z$ and SM particles,
\begin{eqnarray}{\label{SMLag}}
\mathcal{L_{\text{SM}}}&\supset& \frac{h}{\upsilon _{\text{EW}}}\left(\sum_{f} m_{f}\bar{\psi}_{f}\psi_{f}+2m^{2}_{w}W^{+}_{\mu}W^{- \mu}+m^{2}_{z}Z^{\mu}Z_{\mu}\right)
+ Z_{\mu}\sum_{f}\bar{\psi}_{f}\gamma^{\mu}\left(g_{V}-g_{A}\gamma_{5}\right)\psi_{f}+\cdots\nonumber\\
\end{eqnarray}
with
\begin{eqnarray}{\label{gs}}
g_{V}=\frac{g}{\cos\theta_{W}}\left(\frac{T_{3f}}{2}-Q_{f}\sin^{2}\theta_{W}\right)~~~~\text{and}~~~~
g_{A}=\frac{g}{\cos\theta_{W}} \left(\frac{T_{3f}}{2}\right).
\end{eqnarray}
Here, weak scale $\upsilon _{\text{EW}}=246$ GeV,  $g\simeq 0.65$ is the gauge coupling of $SU(2)_{L}$ group, 
$\theta_{W}$ denotes the weak mixing angle,
 and $Q_{f}$ is the electric charge, with $T_{3f}=+(-)\frac{1}{2}$ for up (down)-type SM fermion, respectively. 

Moreover, $\mathcal{L_{\text{dark}}}$ in Eq.(\ref{Lag}) generally contains interactions between Majorana DM (in 4-component notation) and SM mediators \footnote{Note,  there is no vector coupling between Majorana DM and $Z$ boson.},
 \begin{eqnarray}{\label{DarkLag}}
\mathcal{L_{\text{dark}}}(\chi, \cdots)\supset c_{h} h\bar{\chi}\chi + c_{z} Z^{\mu}\bar{\chi}\gamma_{\mu}\gamma_{5}\chi+\cdots,
\end{eqnarray}
The $c_h$-  and $c_z$- interaction term constitute the minimal framework from the perspective of effective field theory, 
where higher-dimensional operators \cite{1407.1859} responsible for obvious gauge invariance of Eq.(\ref{DarkLag}) should be taken into account.
We refer $\mathcal{L_{\text{dark}}}$ in Eq.(\ref{DarkLag}) as  the ``minimal" effective field theory. 

New physical particles beyond the minimal effective field theory impose diverse effects.
If they are decoupled, their net effects are recorded in parameter $c_{h}$ and $c_{z}$ in Eq.(\ref{DarkLag}).
Conversely, if not, they should be included in Eq.(\ref{SMLag}) or Eq.(\ref{DarkLag}),
which either play the role of new mediator between DM and SM sectors or contribute to new DM annihilation final states as long as they are kinetically allowed.
In the former case, Lagrangians in Eq.(\ref{SMLag}) and Eq.(\ref{DarkLag}) contain all possible contributions to DM annihilation and DM scattering cross sections.
In the later one, new particles with a mass of order the weak scale yield a few new Feynman diagrams for these cross sections.
When the number of new particles is large, the numerical treatment is more viable than an analytic one. 
Nevertheless, it is only the analytic treatment which can clearly show us 
the ingredients as required to fit future signatures of DM direct detection, which is the main motivation for this study.

The rest of the paper is organized as follows.
Sec.II is devoted to an analytical derivation of DM annihilation into SM final states in the minimal framework.
We will compare our results with numerical calculation. 
In Sec.III we show the parameter space subject to the latest DM direct detection limits in a model-independent way.
In Sec.IV we apply our method to the singlet-doublet and the minimal supersymmetric standard model (MSSM). 
Finally, we conclude in Sec.V.

\begin{figure}
\includegraphics[width=0.6\textwidth]{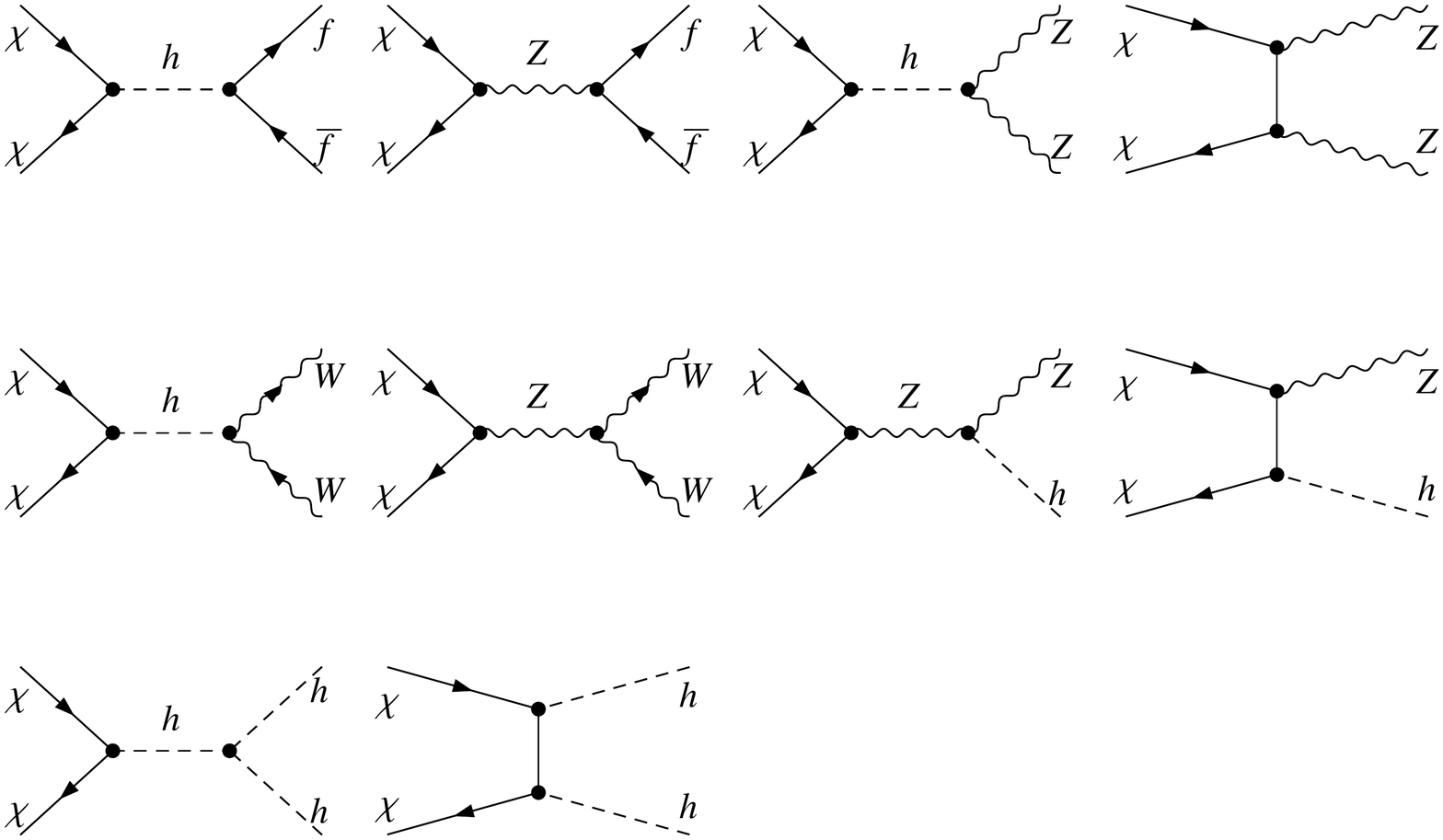}
\vspace{-0.45cm}
\caption{Feynman diagrams for DM annihilation into SM final states.}
\label{Diagram}
\end{figure}

\begin{table}
\begin{center}
\begin{tabular}{ccc}
\hline\hline
${\rm channel}$    & $a$ & $b$   \\  \hline
$f\bar{f}$  &$a_{ff}$  &$b_{ff}$ \\
$ZZ$  &$a_{zz}$ &$b_{zz}$  \\
$W^{+}W^{-}$ &$-$ &$b_{\text{ww}}$\\
$hh$  &$-$ &$b_{hh}$ \\
$Zh$  &$a_{zh}$ &$b_{zh}$ \\
\hline \hline
\end{tabular}
\caption{Coefficients of $\sigma v_{\chi}$ expansion in individual SM final state.}
\end{center}\label{coefficient}
\end{table}

\section{Relic density}
According to the effective Lagrangian in Eq.(\ref{Lag}), 
DM can annihilate into the SM final states such as $f\bar{f}$, $\text{ZZ}$, $\text{WW}$, $Zh$ and $hh$ through SM mediator $h$ and/or $Z$. 
In order to calculate DM relic density,
we firstly derive the thermally averaged cross section $\left<\sigma v_{\chi}\right>$.
The Feynman diagrams responsible for it are shown in Fig.\ref{Diagram}.
Although Feynman diagrams similar to Fig.\ref{Diagram} have already been discussed in more complicated context such as neutralino DM \cite{9207234}, 
a concrete analytic expression for DM annihilation cross section is only viable in some simplified situations 
such as the minimal framework discussed here.

As is well known, the DM annihilation cross section times DM relative velocity $v_{\chi}$ can be expanded in the standard way,
\begin{eqnarray}
\sigma v_{\chi} = a + bv_{\chi}^2 + \mathcal{O}(v_{\chi}^4).
\end{eqnarray}
In Table.\ref{coefficient} we introduce coefficient $a$ and $b$ due to various SM final states in Fig.\ref{Diagram}.
Direct evaluation of them yields
\begin{eqnarray}
{\label{aff}}
a_{ff} &=&\frac{2N_{c}c_{z}^2r_{\text{f$\chi $}} g_{A}^2 m_{f}^2}{\pi m_{Z}^4}, \\
{\label{azz}}
a_{zz} &=& \frac{4c_{z}^4 r_{\text{z$\chi $}} \left(m_{\chi }^2-m_{Z}^2\right)}{\pi  \left(m_{Z}^2-2 m_{\chi }^2\right){}^2}, \\
{\label{azh}}
a_{zh} &=& \frac{c_{z}^2 r_{\chi z h}^3 m_{\chi }^2 }{64\pi  \upsilon _{\text{EW}}^2 m_{Z}^2},
\end{eqnarray}
and
\begin{small}
\begin{eqnarray}
{\label{bff}}
b_{ff} &=& \frac{N_{c}c_{h}^2 m_{f}^2 r_{\text{f$\chi $}} \left(m_{\chi }^2-m_{f}^2\right)}{2 \pi  \upsilon _{\text{EW}}^2 \left(m_{h}^2-4 m_{\chi }^2\right){}^2}
\nonumber\\
&+&\frac{N_{c}c_{z}^2 g_{A}^2 m_{f}^2 \left(5 m_{f}^2-4 m_{\chi }^2\right)}{4 \pi  r_{\text{f$\chi $}} m_{\chi }^2 m_{Z}^4} +   \frac{N_{c}c_{z}^2 r_{\text{f$\chi $}} \left(m^{2}_{f}(g^{2}_{V}-2g^{2}_{A})+2m^{2}_{\chi}(g^{2}_{V}+g^{2}_{A})\right)}{3 \pi \left(m_{Z}^2-4 m_{\chi }^2\right){}^2}, \\
{\label{bww}}
b_{\text{ww}} &=& \frac{c_{h}^2 r_{\text{w$\chi $}} \left(-4 m_{W}^2 m_{\chi }^2+4 m_{\chi }^4+3 m_{W}^4\right)}{4 \pi  \upsilon _{\text{EW}}^2 \left(m_{h}^2-4 m_{\chi }^2\right){}^2}\nonumber\\
&+&\frac{c_{z}^2r_{\text{w$\chi $}} g^2\cos^2\theta_{W}\left(-17 m_{W}^4 m_{\chi }^2+16 m_{\chi }^4m^{2}_{w}-3 m_{W}^6+4m^{6}_{\chi}\right)}{6 \pi  m_{W}^4 \left(m_{Z}^2-4 m_{\chi }^2\right){}^2}, \\
{\label{bzz}}
b_{zz} &=& \frac{c_{h}^2 r_{\text{z$\chi $}}\left(-4 m_{Z}^2 m_{\chi }^2+4 m_{\chi }^4+3 m_{Z}^4\right)}{8\pi \upsilon _{\text{EW}}^2  \left(m_{h}^2-4 m_{\chi }^2\right){}^2}\nonumber\\
&+&\frac{2c_{h} c_{z}^2 r_{\text{z$\chi $}}m_{\chi}\left(-9 m_{Z}^4 m_{\chi }^2+12 m_{Z}^2 m_{\chi }^4-8 m_{\chi }^6+2 m_{Z}^6\right)}{3 \pi \upsilon_{\text{EW}} \left(m_{h}^2-4 m_{\chi }^2\right) \left(m_{Z}^3-2 m_{Z} m_{\chi }^2\right){}^2} \nonumber \\
&+& \frac{c_{z}^4  r_{\text{z$\chi $}}  \left(-118 m_{Z}^8 m_{\chi }^2+172 m_{Z}^6 m_{\chi }^4+32 m_{Z}^4 m_{\chi }^6-192 m_{Z}^2 m_{\chi }^8+128 m_{\chi }^{10}+23 m_{Z}^{10}\right)}{6 \pi \left(m_{Z}^3-2 m_{Z} m_{\chi }^2\right){}^4}, \\
{\label{bhh}}
b_{hh} &=&\frac{9 c_{h}^2 m_{h}^4 r_{\text{h$\chi $}}}{32 \pi  \upsilon _{\text{EW}}^2 \left(m_{h}^2-4 m_{\chi }^2\right){}^2}\nonumber\\
& +&\frac{c_{h}^3 m_{h}^2 r_{\text{h$\chi $}} m_{\chi } \left(2 m_{h}^2-5 m_{\chi }^2\right)}{2 \pi  \upsilon _{\text{EW}} \left(m_{h}^2-4 m_{\chi }^2\right) \left(m_{h}^2-2 m_{\chi }^2\right){}^2}
+\frac{2c_{h}^4 r_{\text{h$\chi $}} m_{\chi }^2 \left(-8 m_{h}^2 m_{\chi }^2+9 m_{\chi }^4+2 m_{h}^4\right)}{3 \pi  \left(m_{h}^2-2 m_{\chi }^2\right){}^4},\\
{\label{bzh}}
b_{zh}&=&\frac{c_{z}^2r_{\chi zh}}{768 \pi \upsilon_{\text{EW}}^2 m_{\chi}^2 m_{Z}^2 (m_{Z}^2-4 m_{\chi}^2)^{2}}  \left[(4 m_{Z}^6 (5 m_{h}^2 + 59 m_{\chi}^2)-2 m_{Z}^4 (5 m_{h}^4 + 74 m_{h}^2 m_{\chi}^2 + 344 m_{\chi}^4) \nonumber \right.\\
 &+& \left.\ 96 m_{\chi}^2 m_{Z}^2 (m_{h}^4 - m_{h}^2 m_{\chi}^2 + 14 m_{\chi}^4)-192 m_{\chi}^4 (m_{h}^4 - 5 m_{h}^2 m_{\chi}^2 + 4 m_{\chi}^4)-10 m_{Z}^8)\right]\nonumber\\
 &+&\frac{c_{h}c_{z}^2 r_{\chi zh}}{12\pi \upsilon_{\text{EW}}m_{\chi} m^{2}_{Z}(4 m_{\chi}^2 - m_{Z}^2) (m_{h}^2 - 4 m_{\chi}^2 + m_{Z}^2)^2}\left[(2 m_{Z}^6 (m_{h}^2 -9 m_{\chi}^2) - 2 m_{\chi}^2 (m_{h}^2 - 4 m_{\chi}^2)^3 \nonumber\right.\\
 &-& \left.\ m_{Z}^4 (m_{h}^4 + 14 m_{h}^2 m_{\chi}^2 - 104 m_{\chi}^4) + 2 m_{\chi}^2 m_{Z}^2 (m_{h}^4 + 8 m_{h}^2 m_{\chi}^2 - 48 m_{\chi}^4) -m_{Z}^8)\right]\nonumber\\
 &+&\frac{c_{h}^2c_{z}^2 r_{\chi zh}}{768 \pi m_{\chi}^2 m_{Z}^2 (m_{h}^2 - 4 m_{\chi}^2 + m_{Z}^2)^4} \left[m_{Z}^{10}+2 m_{Z}^6 (3 m_{h}^4 + 16 m_{\chi}^4) +4 m_{Z}^8 (m_{\chi}^2 - m_{h}^2)\nonumber\right.\\
 &-& \left.\ 4m_{Z}^4 (m_{h}^2 - 4 m_{\chi}^2)^2 (m_{h}^2 + 10 m_{\chi}^2) + 4 m_{\chi}^2 (m_{h}^2 - 4 m_{\chi}^2)^4 \nonumber\right.\\
 &+& \left.\  m_{Z}^2 (m_{h}^2 - 4 m_{\chi}^2)^2 (m_{h}^4 + 8 m_{h}^2 m_{\chi}^2 + 80 m_{\chi}^4) \right]
\end{eqnarray}
\end{small}
where $N_{c}=1(3)$ for SM lepton (quark) and $m_{\chi}$ refers to DM mass.
Functional $r_{ij}$ and $r_{\chi ij}$ is defined as
\begin{eqnarray}
r_{ij}&=&\sqrt{1-m_i^2/m_j^2},\nonumber\\
r_{\chi ij}&=&\sqrt{m_i^4-2 m_i^2 (m_j^2+4 m_{\chi}^2 )+(m_j^2-4 m_{\chi}^2)^2}/m_{\chi}^{2} \nonumber
\end{eqnarray}
, respectively.
 
A few comments are in order regarding our results.
At first, in the case $c_{h}\rightarrow 0$, both $a_{ff}$ and $a_{zz}$ in Eq.(\ref{aff})- Eq.(\ref{azz}) coincide with results of $Z$ portal \cite{1708.00890,1506.06767},
but $b_{ff}$ and $b_{zz}$ in \cite{1708.00890} are both two times of that in Eq.(\ref{bff}) and Eq.(\ref{bzz}), respectively.
Secondly, in the case $c_{z}\rightarrow 0$ all $a$s in Eq.(\ref{aff})- Eq.(\ref{azh}) disappear as the same as in the Higgs portal, 
and our $b_{ff}$ in Eq.(\ref{bff}) and $b_{zz}$ and $b_{hh}$ (the $c^{2}_{h}$-term) is in agreement with the results of \cite{1203.2064} and \cite{0611069}, respectively.
Thirdly, when both $c_{z}$ and $c_{h}$ are non-zero, 
interference effects occur in $b_{zz}$ and $b_{zh}$, which are explicitly shown.
These interference effects can be neglected except in some particular DM mass range between $m_z$ and $m_h$,
where it is not small relative to the other contributions.
Finally, we have also included the SM Higgs self interaction contribution to $b_{hh}$ in Eq.(\ref{bhh}).
We verified that our results agree with the numerical calculation in terms of code MicrOMEGAs \cite{1407.6129}, 
with at most $10\%-15\%$ deviation in the estimate of DM relic density.

\section{Direct~Detection}
The interactions in Eq.(\ref{DarkLag}) yield both spin-dependent (SD) and spin-independent (SI) effective couplings between DM and SM nucleons.
In particular, Yukawa coupling constant $c_{h}$ and $c_{z}$ controls SI and SD scattering cross section, respectively,
which are given by \cite{Jungman:1995df,1605.08442},
\begin{eqnarray}
\sigma_{\text{SI}}&\simeq&c^{2}_{h}\times (2.11\times 10^{3}\text{zb}),\nonumber\\
\sigma^{p}_{\text{SD}}&\simeq& c^{2}_{z}\times (1.17\times 10^{9}\text{zb}),\nonumber\\
\sigma^{n}_{\text{SD}}&\simeq& c^{2}_{z}\times (8.97\times 10^{8}\text{zb}).
\end{eqnarray}
Here nuclear form factors have been chosen as in \cite{0309149}. 
The approximations to $\sigma_{\text{SI}}$ and $\sigma_{\text{SD}}$ are always valid for DM mass 
 $m_{\chi}$ above $\text{a~few~times}$ of $m_{p,n}$.

\begin{figure}
\includegraphics[width=0.6\textwidth]{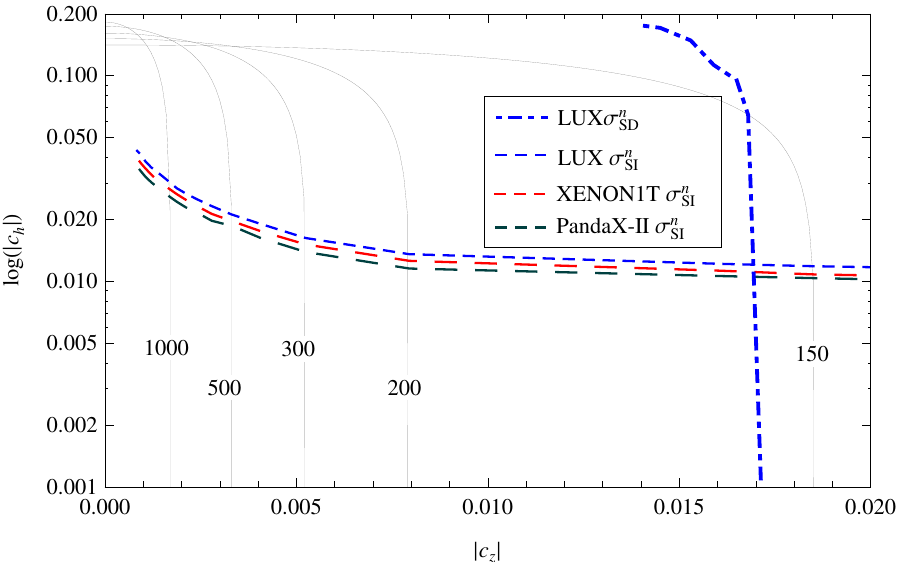}
\vspace{-0.5cm}
\caption{Parameter space of DM relic density in the two-parameter plane of $c_{h}$ and $c_{z}$ 
subject to the latest PandaX-II \cite{1708.06917} (green), Xenon-1T \cite{1705.06655} (red), and LUX 2016 \cite{1608.07648} (blue) limit.
The DM masses (in unit of GeV) referring to each contour are drawn for clarity,
which implies that the model-independent exclusion limit for DM mass is about $\sim 155$ GeV.}
\label{ps}
\end{figure}

In Fig.\ref{ps} we show the parameter space of DM relic density $ \Omega_{\text{DM}}h^{2}=0.1199\pm 0.0027$  \cite{1303.5076} 
in the two-parameter plane of $c_{h}$ and $c_{z}$, with contours referring to DM masses in unit of GeV.
We also draw contours of the latest PandaX-II \cite{1708.06917}, Xenon-1T \cite{1705.06655} and LUX 2016 \cite{1608.07648} limit simultaneously.
Parameter regions above the color lines or on the right hand side of the blue line are excluded,
from which we find that model-independent exclusion limit for DM mass is about $\sim 155$ GeV.
Only a small region
\begin{eqnarray}{\label{parameter}}
0 \leq\mid c_{z}\mid \leq 0.018,~~~~~~0 \leq \mid c_{h}\mid \leq 0.06
\end{eqnarray}
is left for future tests. 
If this region is excluded by future experimental limits,
we can draw the conclusion that 
either new particle(s) is required to appear at the weak scale or simplified Majorana DM models are disfavored.
In what follows we will discuss implication of our results in a few simplified models.

The parameter space of coupling $c_z$ and $c_h$ (alternatively DM mass range) as given by Eq.(\ref{parameter}) is not affected by other constraints such as mono-jet limit \cite{monojet,1408.3583,1604.07773} at LHC or the constraint on DM annihilation cross section $\sigma(\chi\chi\rightarrow \gamma\gamma)$ at Fermi-LAT \cite{1305.5597,1506.00013,1503.02641}.
The mono-jet constraint is sensitive to parameter $c_z$ only for DM mass $m_{\chi}<M_{Z}/2$ in our situation,
which excludes DM mass below $\sim 50$ GeV for $c_{z}=1.0$, see, e.g. \cite{1604.07773}.
It implies that the surviving DM mass range referring to Eq.(\ref{parameter}) is not sensitive to present mono-jet limit. 
On the other hand, the Fermi-LAT constraint on $\gamma$ spectrum is sensitive to both $c_z$ and $c_h$ for DM mass below $\sim 500$ GeV,
where the Feynman diagram for $\sigma(\chi\chi\rightarrow h/Z \rightarrow\gamma\gamma)$  is dominated by top, bottom fermion loop and $W$ boson loop.  As expected, there is a peak in the $\gamma$ spectrum that appears at DM mass close to a half of the mediator mass, i.e, $M_{h}/2$ or $M_{Z}/2$ in our case. When the DM mass such as what corresponds to Eq.(\ref{parameter}) obviously deviates from the pole masses above, 
the Fermi-LAT constraint is weak as well.

 \section{Applications}
 \subsection{Singlet-Doublet Dark Matter}
This model contains two fermion doublets $L'=(l'^{0},l^{-})^{T}$, $L=(l^{+}, l^{0})^{T}$ and a fermion singlet $\psi_{s}$. 
The dark sector Lagrangian $\mathcal{L_{\text{dark}}}$ reads as \cite{0510064, 0705.4493, 0706.0918},
\begin{eqnarray}
\mathcal{L_{\text{dark}}}&=&\frac{i}{2}\left(\bar{\psi_{s}}\sigma^{\mu}\partial_{\mu}\psi_{s}+\bar{L'}\sigma^{\mu}\partial_{\mu}L'+\bar{L}\sigma^{\mu}\partial_{\mu}L\right)
\nonumber\\
&+&\left(-y_{1}L'H\psi_{s}-y_{2}\bar{L}\bar{\psi}_{s}H+\text{H.c}\right)-\frac{m_{\psi_{s}}}{2}\psi_{s}\psi_{s}-m_{D}L'L
\end{eqnarray}
where $m_{s}$, $m_{D}$ and  $y_{1,2}$ are mass and Yukawa coupling parameters, respectively.  
$H$ denotes the SM Higgs doublet. 
In the basis $(\psi_{s}, l'^{0},l^{0})$ the symmetric mass matrix for neutral fermions is given by,
\begin{eqnarray}{\label{Mass1}}
M_{\chi}=\left(
\begin{array}{ccc}
m_{s} & \frac{y_{1}\upsilon_{\text{EW}}}{\sqrt{2}} &  \frac{y_{2}\upsilon_{\text{EW}}}{\sqrt{2}} \\
 *&   0 & m_{D}  \\
* & * & 0  
\end{array}%
\right).
\end{eqnarray}

\begin{figure}
\includegraphics[width=0.6\textwidth]{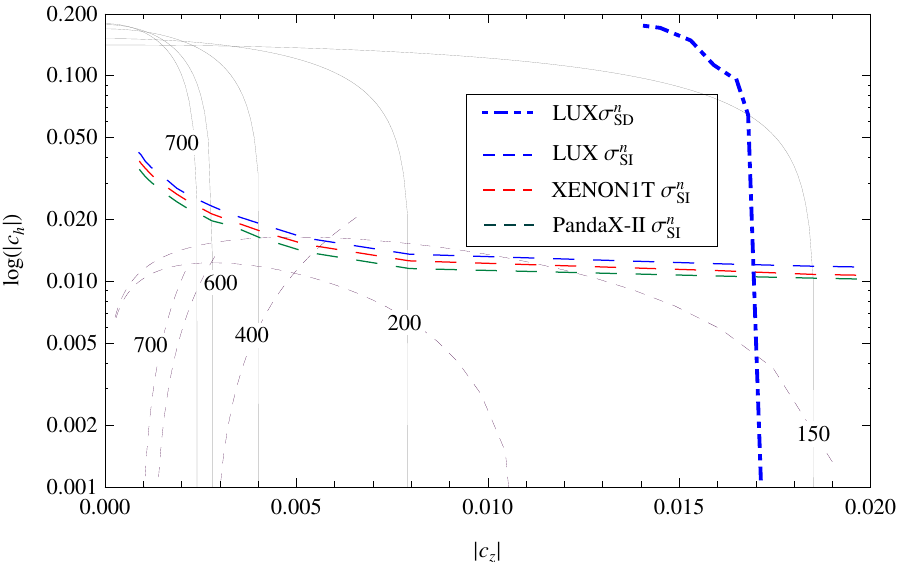}
\vspace{-0.5cm}
\caption{Contours of DM mass in dotted lines projected to the plane of $c_{h}-c_{z}$ for $y_{1}=-3$ and $y_{2}=0.1$. 
Contours of DM relic density and color lines are the same as in Fig.\ref{ps}.
We have imposed the condition that DM mass $m_{\chi}$  should be at least an order of magnitude smaller than $m_{D}$.  }
\label{sd}
\end{figure}

This model is similar to the neutralino sector of the next-to-minimal supersymmetric model (NMSSM) 
when bino and wino components are both decoupled.
Imposing the decoupling limit $m_{D}>> m_{s}, \upsilon_{\text{EW}}$ on the dark sector 
yields only a light singlet-like DM with mass $m_{\chi}\simeq m_{s}$.
Under this limit, the effective coupling $c_{h\bar{\chi}\chi}$ and $c_{z\bar{\chi}\chi}$ reduces to, respectively \cite{1505.03867} ,
 \begin{eqnarray}{\label{sdm}}
c_{h\bar{\chi}\chi}&\simeq&- \frac{\upsilon_{\text{EW}}}{m_{D}}\left(2y_{1}y_{2}+(y^{2}_{1}+y^{2}_{2})\frac{m_{\chi}}{m_{D}}\right),\nonumber\\
c_{z\bar{\chi}\chi}&\simeq& \frac{1}{2}\frac{\upsilon_{\text{EW}}}{m_{D}}\frac{m_{Z}}{m_{D}}(y^{2}_{1}-y^{2}_{2})\left(1-\frac{m^{2}_{\chi}}{m^{2}_{D}}\right).
\end{eqnarray}
Note that $\mid c_{z} \mid$ and $\mid c_{h}\mid$ are both unchanged under the exchange of $y_{1}\leftrightarrow y_{2}$.
Since the parameter ranges in Eq.(\ref{parameter}) favor larger value of $\mid c_{h} \mid$ relative to $\mid c_{z}\mid$,
it implies that the product $y_{1}y_{2}$ in Eq.(\ref{sdm}) should be at most of order $m_{\chi}/m_{D}$.
Otherwise, $\mid c_{h}\mid$ at the crossing points with contours of DM relic density  would be too large
to excess the direct detection limits as shown in Fig.\ref{ps}.

In Fig.\ref{sd} we show the contours of DM mass projected to the plane of $c_{h}-c_{z}$ for $y_{1}=-3$ and $y_{2}=0.1$,
where the condition that DM mass $m_{\chi}\simeq m_{s}$ should be at least an order of magnitude smaller than $m_{D}$ has been imposed. 
The crossing points with the contours of DM relic density are indeed beneath the DM direct detection limits for DM mass range between $200$ GeV and $600$ GeV. When the magnitude of $y_{1}$ is tuned to be smaller than 2, these viable crossing points disappear.

\subsection{MSSM}
Now, we discuss application to MSSM with decoupling mass spectrum, 
in which all supersymmetric particles except the lightest neutralino are decoupled from the weak scale. 
The symmetric neutralino mass matrix $M_{\chi}$ under the gauge eigenstates $(\tilde{B}^{0}, \tilde{W}^{0},\tilde{H}^{0}_{d},\tilde{H^{0}_{u}})$ is given by,
\begin{eqnarray}{\label{Mass2}}
M_{\chi}=\left(
\begin{array}{cccc}
M_{1} & 0 &  -m_{Z}s_{W}\cos\beta&m_{Z}s_{W}\sin\beta  \\
 *&   M_{2}  & m_{Z}c_{W}\cos\beta  &  -m_{Z}c_{W}\sin\beta  \\
* & * & 0 & -\mu  \\
* & * & * & 0  \\
\end{array}%
\right).
\end{eqnarray}
Imposing the decoupling limit on the Higgs sector and the neutralino sector by $\mid\mu\mid, M_{1}>> M_{2}, m_{Z}$ simultaneously 
leads to a wino-like DM with mass $m_{\chi^{0}_{1}}\simeq M_{2}$ and reduced effective coupling coefficient $c_{h}$ and $c_{z}$ \cite{0505142,1512.02472, 1506.07177},
\begin{eqnarray}{\label{cz1}}
c_{h\bar{\chi}\chi}&\simeq&\frac{g}{4}\cos\theta_{W}\frac{m_{Z}}{\mu}\left(\frac{m_{\chi^{0}_{1}}}{\mu}+\sin2\beta\right),\nonumber\\
c_{z\bar{\chi}\chi}&\simeq&-\frac{g}{4}\cos\theta_{W} \frac{m^{2}_{Z}}{\mu^{2}}\left(1-\frac{m^{2}_{\chi^{0}_{1}}}{\mu^{2}}\right),
\end{eqnarray}
respectively.
Instead, imposing a different decoupling limit $\mid\mu\mid, M_{2}>>M_{1}, m_{Z}$ we obtain a bino-like DM with mass $m_{\chi^{0}_{1}}\simeq M_{1}$ and 
\begin{eqnarray}{\label{cz2}}
c_{h\bar{\chi}\chi}&\simeq&\frac{g}{4}\sin\theta_{W}\frac{m_{Z}}{\mu}\left(\frac{m_{\chi^{0}_{1}}}{\mu}+\sin2\beta\right),\nonumber\\
c_{z\bar{\chi}\chi}&\simeq&-\frac{g}{4}\sin\theta_{W} \frac{m^{2}_{Z}}{\mu^{2}}\left(1-\frac{m^{2}_{\chi^{0}_{1}}}{\mu^{2}}\right).
\end{eqnarray}

Both decoupling limits yield a light chargino $\tilde{\chi}^{\pm}$ with mass slightly larger than DM mass.
For bino-like DM the modification to DM annihilation cross section can be ignored,
whereas for wino-like DM, the correction due to chargino-exchanging DM annihilation into $W^{+}W^{-}$ is small 
\footnote{Although kinetically allowed, the ratio between s-wave contribution to DM annihilation due to $W^{+}W^{-}$ and top quark final states is of order $\sim m^{4}_{Z}/m^{2}_{\chi}m^{2}_{t}$, where $m_{t}$ is the top quark mass.
Thus,  the s-wave of chargino-exchanging $W$ bosons final states is subdominant 
in compared with that of top quark final states in DM mass range $m_{\chi}>m_{t}$.}, 
apart from a large contribution due to co-annihilation which occurs in DM mass range above $\sim 1$ TeV \cite{0610249}.
Under the decoupling limit, $\mid c_{z} \mid < 1.0\times 10^{-3}$ in Eq.(\ref{cz1})-Eq.(\ref{cz2}),
given $\mid m_{\chi^{0}_{1}}/\mu \mid \leq 0.1$ and $m_{Z}/\mid \mu \mid \leq 0.1$.
From the contours of DM relic density in Fig.\ref{ps},  one finds that wino-like DM with DM mass below $1$ TeV is excluded,
which is consistent with the concrete estimate of wino-like DM mass in Ref.\cite{0610249}.

\section{Conclusion}
In this paper, we have revisited the Majorana DM, a weakly interacting massive particle, from the viewpoint of minimal effective field theory.
Unlike the Dirac-type analogy, there is no vector coupling between Majorana DM and the $Z$ boson.
In this framework, there are only three parameters, i.e., the DM mass and Yukawa coupling constant $c_{h}$ and $c_{Z}$.
Accordingly, it is sufficient to constrain the parameter space in relatively model independent way. 
In order to achieve this, an analytical derivation of DM annihilation into all possible SM final states was performed, 
which included contributions such as interference effects and SM Higgs self-interaction as well.
The fit to the latest LUX, PandaX-II and Xenon-1T limits points to DM mass lower bound about $\sim 155$ GeV.
Also, preliminary applications to singlet-doublet and MSSM have been addressed. 
In singlet-doublet model, 
we found that singlet-like DM with mass range between $200$ GeV and $600$ GeV still survives in the latest DM direct detection limits.
In the MSSM with decoupled mass spectrum, 
we recovered exclusion limits on neutralino DM mass such as wino-like DM.

\begin{acknowledgments}
The authors are grateful to referees for useful suggestions. 
This work is supported in part by the National Natural Science Foundation of China under Grant No.11775039
and Postdoctoral Science Foundation of China under Grant No. 2017M611008.
\end{acknowledgments}

\end{document}